\documentclass[11pt]{article}
\usepackage{DGfest,epsfig}
\usepackage{times}

\def\be{\begin{equation}}
\def\ee{\end{equation}}
\def\bea{\begin{eqnarray}}
\def\eea{\end{eqnarray}}

\def\Tr{{\rm Tr}}
\def\re{{\rm Re}}
\def\im{{\rm Im}}

\begin{document}
\baselineskip 11.5pt
\title{DERIVING THE POMERON FROM A EUCLIDEAN-MINKOWSKIAN DUALITY}

\author{ E. MEGGIOLARO }

\address{Dipartimento di Fisica, Universit\`a di Pisa,  and I.N.F.N.,
Sezione di Pisa,\\ Largo Pontecorvo 3, I--56127 Pisa, Italy}

\maketitle\abstracts{
After a brief review, in the first part, of some relevant analyticity
properties of the loop--loop scattering amplitudes in gauge theories, when
going from Minkowskian to Euclidean theory, in the second part we shall see
how they can be related to the still unsolved problem of the s--dependence
of the hadron--hadron total cross--sections.}

\section{Loop--loop scattering amplitudes}

Differently from the parton--parton scattering amplitudes, which are known to
be affected by infrared (IR) divergences, the elastic scattering amplitude of
two colourless states in gauge theories, e.g., two $q \bar{q}$ meson states,
is expected to be an IR--finite physical quantity.
It was shown in Refs. \cite{Nachtmann97,Dosch,Berger} that the high--energy
meson--meson elastic scattering amplitude can be approximately reconstructed
by first evaluating, in the eikonal approximation, the elastic scattering
amplitude of two $q \bar{q}$ pairs (usually called ``{\it dipoles}''), of
given transverse sizes $\vec{R}_{1\perp}$ and $\vec{R}_{2\perp}$ respectively,
and then averaging this amplitude over all possible values of
$\vec{R}_{1\perp}$ and $\vec{R}_{2\perp}$ with two proper squared
wave functions $|\psi_1 (\vec{R}_{1\perp})|^2$ and
$|\psi_2 (\vec{R}_{2\perp})|^2$, describing the two interacting mesons.
The high--energy elastic scattering amplitude of two {\it dipoles} is
governed by the (properly normalized) correlation function of two Wilson loops
${\cal W}_1$ and ${\cal W}_2$, which follow the classical straight lines for
quark (antiquark) trajectories:
\be
{\cal M}_{(ll)} (s,t;~\vec{R}_{1\perp},\vec{R}_{2\perp}) \equiv
-i~2s \displaystyle\int d^2 \vec{z}_\perp
e^{i \vec{q}_\perp \cdot \vec{z}_\perp}
\left[ {\langle {\cal W}_1 {\cal W}_2 \rangle \over
\langle {\cal W}_1 \rangle \langle {\cal W}_2 \rangle} -1 \right] ,
\label{scatt-loop}
\ee
where $s$ and $t = -|\vec{q}_\perp|^2$ ($\vec{q}_\perp$ being the transferred
momentum) are the usual Mandelstam variables.
More explicitly the Wilson loops ${\cal W}_1$ and ${\cal W}_2$ are so defined:
\bea
{\cal W}^{(T)}_1 &\equiv&
{1 \over N_c} \Tr \left\{ {\cal P} \exp
\left[ -ig \displaystyle\oint_{{\cal C}_1} A_\mu(x) dx^\mu \right] \right\} ,
\nonumber \\
{\cal W}^{(T)}_2 &\equiv&
{1 \over N_c} \Tr \left\{ {\cal P} \exp
\left[ -ig \displaystyle\oint_{{\cal C}_2} A_\mu(x) dx^\mu \right] \right\} ,
\label{QCDloops}
\eea
where ${\cal P}$ denotes the ``{\it path ordering}'' along the given path
${\cal C}$ and $A_\mu = A_\mu^a T^a$;
${\cal C}_1$ and ${\cal C}_2$ are two rectangular paths which
follow the classical straight lines for quark [$X_{(+)}(\tau)$, forward in
proper time $\tau$] and antiquark [$X_{(-)}(\tau)$, backward in $\tau$]
trajectories, i.e.,
\bea
{\cal C}_1 &\to&
X_{(\pm 1)}^\mu(\tau) = z^\mu + {p_1^\mu \over m} \tau
\pm {R_1^\mu \over 2} , \nonumber \\
{\cal C}_2 &\to&
X_{(\pm 2)}^\mu(\tau) = {p_2^\mu \over m} \tau \pm {R_2^\mu \over 2} ,
\label{traj}
\eea
and are closed by straight--line paths at proper times $\tau = \pm T$, where
$T$ plays the role of an IR cutoff \cite{Verlinde,Meggiolaro02}, which must
be removed at the end ($T \to \infty$).
Here $p_1$ and $p_2$ are the four--momenta of the two quarks and of the two
antiquarks with mass $m$, moving with speed $\beta$ and $-\beta$ along, for
example, the $x^1$--direction:
\bea
p_1 &=& m (\cosh {\chi \over 2},\sinh {\chi \over 2},0,0) ,
\nonumber \\
p_2 &=& m (\cosh {\chi \over 2},-\sinh {\chi \over 2},0,0) ,
\label{p1p2}
\eea
where $\chi = 2~{\rm arctanh} \beta > 0$ is the hyperbolic angle between the
two trajectories $(+1)$ and $(+2)$.
Moreover, $R_1 = (0,0,\vec{R}_{1\perp})$, $R_2 = (0,0,\vec{R}_{2\perp})$
and $z = (0,0,\vec{z}_\perp)$, where $\vec{z}_\perp = (z^2,z^3)$ is the
impact--parameter distance between the two loops in the transverse plane.

The expectation values $\langle {\cal W}_1 {\cal W}_2 \rangle$,
$\langle {\cal W}_1 \rangle$, $\langle {\cal W}_2 \rangle$ are averages
in the sense of the QCD functional integrals:
\be
\langle {\cal O}[A] \rangle =
{1 \over Z} \displaystyle\int [dA] \det(Q[A]) e^{iS_A} {\cal O}[A] ,
\ee
where $Z = \displaystyle\int [dA] \det(Q[A]) e^{iS_A}$, $S_A$ is the
pure--gauge (Yang--Mills) action and $Q[A]$ is the {\it quark matrix}.

It is convenient to consider also
the correlation function of two Euclidean Wilson loops
$\tilde{\cal W}_1$ and $\tilde{\cal W}_2$ running along two rectangular paths
$\tilde{\cal C}_1$ and $\tilde{\cal C}_2$ which follow the following
straight--line trajectories:
\bea
\tilde{\cal C}_1 &\to&
X^{(\pm 1)}_{E\mu}(\tau) = z_{E\mu} + {p_{1E\mu} \over m}
\tau \pm {R_{1E\mu} \over 2} , \nonumber \\
\tilde{\cal C}_2 &\to&
X^{(\pm 2)}_{E\mu}(\tau) = {p_{2E\mu} \over m} \tau
\pm {R_{2E\mu} \over 2} ,
\label{trajE}
\eea
and are closed by straight--line paths at proper times $\tau = \pm T$. Here
$R_{1E} = (0,\vec{R}_{1\perp},0)$, $R_{2E} = (0,\vec{R}_{2\perp},0)$ and
$z_E = (0,\vec{z}_\perp,0)$. Moreover, in the Euclidean theory we {\it choose}
the four--vectors $p_{1E}$ and $p_{2E}$ to be:
\bea
p_{1E} &=& m (\sin{\theta \over 2}, 0, 0, \cos{\theta \over 2} ) , \nonumber \\
p_{2E} &=& m (-\sin{\theta \over 2}, 0, 0, \cos{\theta \over 2} ) ,
\label{p1p2E}
\eea
$\theta \in (0,\pi)$ being the angle formed by the two trajectories
$(+1)$ and $(+2)$ in Euclidean four--space.\\
Let us introduce the following notations for the normalized loop--loop
correlators in the Minkowskian and in the Euclidean theory,
in the presence of a {\it finite} IR cutoff $T$:
\bea
{\cal G}_M(\chi;~T;~\vec{z}_\perp,\vec{R}_{1\perp},\vec{R}_{2\perp}) &\equiv&
{ \langle {\cal W}^{(T)}_1 {\cal W}^{(T)}_2 \rangle \over
\langle {\cal W}^{(T)}_1 \rangle
\langle {\cal W}^{(T)}_2 \rangle } ,\nonumber \\ 
{\cal G}_E(\theta;~T;~\vec{z}_\perp,\vec{R}_{1\perp},\vec{R}_{2\perp}) &\equiv&
{ \langle \tilde{\cal W}^{(T)}_1 \tilde{\cal W}^{(T)}_2 \rangle_E \over
\langle \tilde{\cal W}^{(T)}_1 \rangle_E
\langle \tilde{\cal W}^{(T)}_2 \rangle_E } ,
\label{GM-GE}
\eea
where the expectation values $\langle \ldots \rangle_E$ are averages
in the sense of the Euclidean functional integrals:
\bea
\langle {\cal O}[A^{(E)}] \rangle_E &=&
{1 \over Z^{(E)}} \displaystyle\int [dA^{(E)}] \det(Q^{(E)}[A^{(E)}])
e^{-S^{(E)}_A} {\cal O}[A^{(E)}] ,\nonumber \\
Z^{(E)} &=& \displaystyle\int [dA^{(E)}] \det(Q^{(E)}[A^{(E)}]) e^{-S^{(E)}_A} .
\eea
As already stated in Ref. \cite{Meggiolaro02}, the two quantities in Eq.
(\ref{GM-GE}) (with $\chi > 0$ and $0 < \theta < \pi$) are expected to be
connected by the same analytic continuation in the angular variables and in
the IR cutoff which was already derived in the case of Wilson lines
\cite{Meggiolaro02,Meggiolaro97,Meggiolaro98}, i.e.:
\bea
{\cal G}_E(\theta;~T;~\vec{z}_\perp,\vec{R}_{1\perp},\vec{R}_{2\perp}) &=&
{\cal G}_M(\chi \to i\theta;~T \to -iT;
~\vec{z}_\perp,\vec{R}_{1\perp},\vec{R}_{2\perp}) ,
\nonumber \\
{\cal G}_M(\chi;~T;~\vec{z}_\perp,\vec{R}_{1\perp},\vec{R}_{2\perp}) &=&
{\cal G}_E(\theta \to -i\chi;~T \to iT;
~\vec{z}_\perp,\vec{R}_{1\perp},\vec{R}_{2\perp}) .
\label{analytic}
\eea
Indeed it can be proved \cite{Meggiolaro05}, simply by adapting step by step
the proof derived in Ref. \cite{Meggiolaro02} from the case of Wilson lines to
the case of Wilson loops, that the analytic continuation (\ref{analytic}) is an
{\it exact} result, i.e., not restricted to some order in perturbation theory
or to some other approximation, and is valid both for the Abelian and the
non--Abelian case.
This result is derived under the assumption that the function $G_M$, as a
function of the {\it complex} variable $\chi$, is {\it analytic} in a
domain ${\cal D}_M$ which includes the positive real axis $(\re\chi > 0,
\im\chi = 0)$ and the imaginary segment $(\re\chi = 0, 0 < \im\chi < \pi)$;
and, therefore, the function $G_E$, as a
function of the {\it complex} variable $\theta$, is {\it analytic} in a
domain ${\cal D}_E = \{ \theta \in {\bf C} ~|~ i\theta \in {\cal D}_M \}$,
which includes the real segment $(0 < \re\theta < \pi, \im\theta = 0)$ and the
negative imaginary axis $(\re\theta = 0, \im\theta < 0)$.
The validity of this assumption is confirmed by explicit calculations in
perturbation theory \cite{Meggiolaro97,Meggiolaro05,BB}.
Eq. (\ref{analytic}) is then intended to be valid for every
$\chi \in {\cal D}_M$ (i.e., for every $\theta \in {\cal D}_E$).

As we have said above, the loop--loop correlation functions (\ref{GM-GE}),
both in the Minkowskian and in the Euclidean theory, are expected to be
IR--{\it finite} quantities, i.e., to have finite limits when $T \to \infty$,
differently from what happens in the case of Wilson lines.
One can then define the following loop--loop correlation functions
with the IR cutoff removed:
\bea
{\cal C}_M(\chi;~\vec{z}_\perp,\vec{R}_{1\perp},\vec{R}_{2\perp}) &\equiv&
\displaystyle\lim_{T \to \infty} \left[
{\cal G}_M(\chi;~T;~\vec{z}_\perp,\vec{R}_{1\perp},\vec{R}_{2\perp})
- 1 \right] , \nonumber \\
{\cal C}_E(\theta;~\vec{z}_\perp,\vec{R}_{1\perp},\vec{R}_{2\perp})
&\equiv& \displaystyle\lim_{T \to \infty} \left[
{\cal G}_E(\theta;~T;~\vec{z}_\perp,\vec{R}_{1\perp},\vec{R}_{2\perp})
- 1 \right] .
\label{C12}
\eea
It has been proved in Ref. \cite{Meggiolaro05} that, under certain analyticity
hypotheses in the {\it complex} variable $T$ [hypotheses which are also
sufficient to make the relations (\ref{analytic}) meaningful], the two
quantities (\ref{C12}), obtained {\it after} the removal of the IR cutoff
($T \to \infty$), are still connected by the usual analytic continuation in
the angular variables only:
\bea
{\cal C}_E(\theta;~\vec{z}_\perp,\vec{R}_{1\perp},\vec{R}_{2\perp}) &=&
{\cal C}_M(\chi \to i\theta;~\vec{z}_\perp,\vec{R}_{1\perp},\vec{R}_{2\perp}) ,
\nonumber \\
{\cal C}_M(\chi;~\vec{z}_\perp,\vec{R}_{1\perp},\vec{R}_{2\perp}) &=&
{\cal C}_E(\theta \to -i\chi;
~\vec{z}_\perp,\vec{R}_{1\perp},\vec{R}_{2\perp}) .
\label{final}
\eea
This is a highly non--trivial result, whose general validity is discussed
in Ref. \cite{Meggiolaro05}, where it was also explicitly verified in the
simple case of {\it quenched} QED, where vacuum polarization effects, arising
from the presence of loops of dynamical fermions, are neglected, and in the 
case of a non--Abelian gauge theory with $N_c$ colours, up to the order
${\cal O}(g^4)$ in perturbation theory. Indeed, the validity of the relation
(\ref{final}) has been also recently verified in Ref. \cite{BB} by an explicit
calculation up to the order ${\cal O}(g^6)$ in perturbation theory.

As said in Ref. \cite{Meggiolaro05},
if ${\cal G}_M$ and ${\cal G}_E$, considered as functions of the
{\it complex} variable $T$, have in $T=\infty$ an ``eliminable {\it isolated}
singular point'' [i.e., they are analytic functions of $T$ in the {\it complex}
region $|T| > R$, for some $R \in \Re^+$, and the {\it finite} limits
(\ref{C12}) exist when letting the {\it complex} variable $T \to \infty$],
then, of course, the analytic continuation (\ref{final}) immediately derives
from Eq. (\ref{analytic}) (with $|T| > R$), when letting $T \to +\infty$.
(For example, if ${\cal G}_M$ and ${\cal G}_E$ are analytic functions of $T$
in the {\it complex} region $|T| > R$, for some $R \in \Re^+$,
and they are bounded at large $T$, i.e., $\exists B_{M,E} \in \Re^+$ such that
$|{\cal G}_{M,E}(T)| < B_{M,E}$ for $|T| > R$, then $T=\infty$
is an ``eliminable singular point'' for both of them.)
But the same result (\ref{final}) can also be derived under different
conditions. For example, let us assume that ${\cal G}_E$ is a bounded
analytic function of $T$ in the sector $0 \le \arg T \le {\pi \over 2}$,
with finite limits along the two straight lines on the border of the sector:
${\cal G}_E \to G_{E1}$, for $(\re T \to +\infty,~\im T = 0)$, and
${\cal G}_E \to G_{E2}$, for $(\re T = 0,~\im T \to +\infty)$.
And, similarly, let us assume that ${\cal G}_M$ is a bounded
analytic function of $T$ in the sector $-{\pi \over 2} \le \arg T \le 0$,
with finite limits along the two straight lines on the border of the sector:
${\cal G}_M \to G_{M1}$, for $(\re T \to +\infty,~\im T = 0)$, and
${\cal G}_M \to G_{M2}$, for $(\re T = 0,~\im T \to -\infty)$.
We can then apply the ``Phragm\'en--Lindel\"of theorem'' (see, e.g., Theorem
5.64 in Ref. \cite{PLT}) to state that $G_{E2} = G_{E1}$ and
$G_{M2} = G_{M1}$. Therefore, also in this case, the analytic continuation
(\ref{final}) immediately derives from Eq. (\ref{analytic}) when
$T \to \infty$.

\section{How a pomeron--like behaviour can be derived}

The relation (\ref{final}) allows the derivation of the {\it loop--loop
scattering amplitude} (\ref{scatt-loop}), which we rewrite as
\be
{\cal M}_{(ll)} (s,t;~\vec{R}_{1\perp},\vec{R}_{2\perp}) = -i~2s~
\tilde{\cal C}_M (\chi \to \infty, t;~\vec{R}_{1\perp},\vec{R}_{2\perp}) ,
\label{scatt-loop2}
\ee
$\tilde{\cal C}_M$ being the two--dimensional Fourier transform of
${\cal C}_M$, with respect to the impact parameter $\vec{z}_\perp$,
at transferred momentum $\vec{q}_\perp$ (with $t = -\vec{q}_\perp^2$), i.e.,
\be
\tilde{\cal C}_M (\chi, t;~\vec{R}_{1\perp},\vec{R}_{2\perp}) \equiv
\displaystyle\int d^2 \vec{z}_\perp e^{i \vec{q}_\perp \cdot \vec{z}_\perp}
{\cal C}_M (\chi;~\vec{z}_\perp,\vec{R}_{1\perp},\vec{R}_{2\perp}) ,
\label{CMtilde}
\ee
from the analytic continuation $\theta \to -i\chi$ of the corresponding
Euclidean quantity:
\be
\tilde{\cal C}_E (\theta, t;~\vec{R}_{1\perp},\vec{R}_{2\perp}) \equiv
\displaystyle\int d^2 \vec{z}_\perp e^{i \vec{q}_\perp \cdot \vec{z}_\perp}
{\cal C}_E (\theta;~\vec{z}_\perp,\vec{R}_{1\perp},\vec{R}_{2\perp}) ,
\label{CEtilde}
\ee
which can be evaluated non--perturbatively by well--known and well--established
techniques available in the Euclidean theory.\\
We remind that the hadron--hadron elastic scattering amplitude
${\cal M}_{(hh)}$ can be obtained by averaging the loop--loop scattering
amplitude (\ref{scatt-loop2}) over all possible dipole transverse separations
$\vec{R}_{1\perp}$ and $\vec{R}_{2\perp}$ with two proper squared hadron
wave functions:
\be
{\cal M}_{(hh)}(s,t) =
\displaystyle\int d^2\vec{R}_{1\perp} |\psi_1(\vec{R}_{1\perp})|^2
\displaystyle\int d^2\vec{R}_{2\perp} |\psi_2(\vec{R}_{2\perp})|^2
{\cal M}_{(ll)} (s,t;~\vec{R}_{1\perp},\vec{R}_{2\perp}) .
\label{scatt-hadron}
\ee
(For a detailed description of the procedure leading from the loop--loop
scattering amplitude ${\cal M}_{(ll)}$ to the hadron--hadron elastic
scattering amplitude ${\cal M}_{(hh)}$ we refer the reader to Refs.
\cite{Nachtmann97,Dosch,Berger,LLCM1}. See also Ref. \cite{pomeron-book}
and references therein.)\\
Denoting with ${\cal C}_M^{(hh)}$ and ${\cal C}_E^{(hh)}$ the quantities
obtained by averaging the corresponding loop--loop correlation functions
${\cal C}_M$ and ${\cal C}_E$ over all possible dipole transverse separations
$\vec{R}_{1\perp}$ and $\vec{R}_{2\perp}$, in the same sense as in Eq.
(\ref{scatt-hadron}), i.e.,
\bea
{\cal C}_M^{(hh)} (\chi;~\vec{z}_\perp) &\equiv&
\displaystyle\int d^2\vec{R}_{1\perp} |\psi_1(\vec{R}_{1\perp})|^2
\displaystyle\int d^2\vec{R}_{2\perp} |\psi_2(\vec{R}_{2\perp})|^2
\nonumber \\
&\times& {\cal C}_M (\chi;~\vec{z}_\perp,\vec{R}_{1\perp},\vec{R}_{2\perp}) ,
\nonumber \\
{\cal C}_E^{(hh)} (\theta;~\vec{z}_\perp) &\equiv&
\displaystyle\int d^2\vec{R}_{1\perp} |\psi_1(\vec{R}_{1\perp})|^2
\displaystyle\int d^2\vec{R}_{2\perp} |\psi_2(\vec{R}_{2\perp})|^2
\nonumber \\
&\times& {\cal C}_E (\theta;~\vec{z}_\perp,\vec{R}_{1\perp},\vec{R}_{2\perp}) ,
\label{CMEhh}
\eea
we can write:
\be
{\cal M}_{(hh)} (s,t) = -i~2s~
\tilde{\cal C}_M^{(hh)} (\chi \to \infty, t) ,
\label{scatt-hadron2}
\ee
where, as usual:
\bea
\tilde{\cal C}_M^{(hh)} (\chi, t) &\equiv&
\displaystyle\int d^2 \vec{z}_\perp e^{i \vec{q}_\perp \cdot \vec{z}_\perp}
{\cal C}_M^{(hh)} (\chi;~\vec{z}_\perp) , \nonumber \\
\tilde{\cal C}_E^{(hh)} (\theta, t) &\equiv&
\displaystyle\int d^2 \vec{z}_\perp e^{i \vec{q}_\perp \cdot \vec{z}_\perp}
{\cal C}_E^{(hh)} (\theta;~\vec{z}_\perp) .
\label{CMEhh-tilde}
\eea
Clearly, by virtue of the relation (\ref{final}), we also have that:
\be
\tilde{\cal C}_M^{(hh)} (\chi, t) =
\tilde{\cal C}_E^{(hh)} (\theta \to -i\chi, t) .
\label{final-hh}
\ee
We also remind that, in order to obtain the correct
$s$--dependence of the scattering amplitude (\ref{scatt-hadron2}), one must
express the hyperbolic angle $\chi$ between the two loops in terms of $s$,
in the high--energy limit $s \to \infty$ (i.e., $\chi \to \infty$):
\be
\cosh \chi = {s - m_1^2 - m_2^2 \over 2m_1m_2} ~,~~~~{\rm i.e.:}~~
\chi \mathop{\sim}_{s \to \infty} \log \left( {s \over m_1 m_2} \right) ,
\label{logs}
\ee
where $m_1$ and $m_2$ are the masses of the two hadrons considered.\\
This approach has been extensively used in the literature in order to
address, from a theoretical point of view, the still unsolved problem
of the asymptotic $s$--dependence of hadron--hadron elastic scattering
amplitudes and total cross sections.\\
For example, in Ref. \cite{LLCM2} the loop--loop Euclidean correlation
functions have been evaluated in the context of the so--called ``loop--loop
correlation model'' \cite{LLCM1}, in which the QCD vacuum is described by
perturbative gluon exchange and the non--perturbative ``Stochastic Vacuum
Model'' (SVM), and then they have been continued to the corresponding
Minkowskian correlation functions using the above--mentioned analytic
continuation in the angular variables: the result is an $s$--independent
correlation function
$\tilde{\cal C}_M (\chi \to \infty, t;~\vec{R}_{1\perp},\vec{R}_{2\perp})$
and, therefore, a loop--loop scattering amplitude (\ref{scatt-loop2})
linearly rising with $s$. By virtue of the ``optical theorem'',
\be
\sigma_{\rm tot}^{(hh)} (s) \mathop{\sim}_{s \to \infty}
{1 \over s} {\rm Im} {\cal M}_{(hh)} (s, t=0) ,
\label{optical}
\ee
this should imply (apart from possible $s$--dependences in the hadron wave
functions!) $s$--independent hadron--hadron total cross sections in the
asymptotic high--energy limit, in apparent contradiction to the experimental
observations, which seem to be well described by a ``{\it pomeron}--like''
high--energy behaviour (see, for example, Ref. \cite{pomeron-book} and
references therein):
\be
\sigma_{\rm tot}^{(hh)} (s) \mathop{\sim}_{s \to \infty}
\sigma_0^{(hh)} \left( {s \over s_0} \right)^{\epsilon_P} ,
~~~~{\rm with:}~~\epsilon_P \simeq 0.08 .
\label{pomeron}
\ee
In Refs. \cite{Dosch,Berger} a possible $s$--dependence in the hadron wave
functions was advocated in order to reproduce the phenomenological
``{\it pomeron}--like'' high--energy behaviour of the total cross sections.
However, it would be surely preferable to ascribe the {\it universal}
high--energy behaviour of hadron--hadron total cross sections [the only
dependence on the initial--state hadrons is in the multiplicative constant
$\sigma_0^{(hh)}$ in Eq. (\ref{pomeron})] to the same {\it fundamental}
quantity, i.e., the loop--loop scattering amplitude.
(For a different, but still phenomenological, approach in this direction,
using the SVM, see Ref. \cite{LLCM1}.)\\
The same approach, based on the analytic continuation from Euclidean to
Minkowskian correlation functions, has been also adopted in Ref.
\cite{instanton1} in order to study the one--instanton contribution to
both the line--line (see also Ref. \cite{instanton2}) and the loop--loop
scattering amplitudes: one finds that, after the analytic continuation,
the colour--elastic line--line and loop--loop correlation functions decays
as $1/s$ with the energy. (Instead, the colour--changing inelastic line--line
correlation function is of order $s^0$ and dominates at high energy.
In a further paper \cite{instanton3}, instanton--induced inelastic collisions
have been investigated in more detail and shown to produce total cross sections
rising with $s$.)\\
A behaviour like the one of Eq. (\ref{pomeron}) seems to emerge directly
(apart from possible undetermined $\log s$ prefactors) when applying
the Euclidean--to--Minkowskian analytic--continuation approach to the
study of the line--line/loop--loop scattering amplitudes in strongly coupled
(confining) gauge theories using the AdS/CFT correspondence \cite{JP2,Janik}.
(In a previous paper \cite{JP1} the same approach was also used to study
the loop--loop scattering amplitudes in the ${\cal N} = 4$ SYM theory in
the limit of large number of colours, $N_c \to \infty$, and strong coupling.)\\
It has been also recently proved in Ref. \cite{BB}, by an explicit
perturbative calculation, that the loop--loop scattering amplitude approaches,
at sufficiently high energy, the BFKL--{\it pomeron} behaviour \cite{BFKL}.

The way in which a {\it pomeron}--like behaviour can emerge,
using the Euclidean--to--Minkowskian analytic continuation, was first
shown in Ref. \cite{Meggiolaro97} in the case of the line--line (i.e.,
parton--parton) scattering amplitudes and can be easily readapted to the
case of the loop--loop scattering amplitudes.\\
One simply starts by writing the Euclidean hadronic correlation function
in a partial--wave expansion:
\be
\tilde{\cal C}_E^{(hh)} (\theta,t) =
\displaystyle\sum_{l=0}^{\infty} A_l(t) P_l (\cos \theta) ,
\label{pwe}
\ee
which, by making a Sommerfeld--Watson transformation, can be rewritten in the
following way:
\be
\tilde{\cal C}_E^{(hh)} (\theta,t) = {1 \over 2i}
\displaystyle\int_C {A_l(t) P_l(-\cos\theta) \over \sin(\pi l)} dl ,
\label{swt}
\ee
where ``$C$'' is a contour in the complex $l$--plane, running anticlockwise
around the real positive $l$--axis and enclosing all non--negative integers,
while excluding all the singularities of $A_l$. Eq. (\ref{swt}) can be
verified after recognizing that $P_l (-\cos \theta)$ is an integer function
of $l$ and that the singularities enclosed by the contour $C$ of the
expression under integration in the Eq. (\ref{swt}) are simple poles at 
the non--negative integer values of $l$. So the right--hand side of
(\ref{swt}) is equal to the sum of the residues of the integrand in these
poles and this gives exactly the right--hand side of (\ref{pwe}). The
``{\it minus}'' sign in the argument of the Legendre function $P_l$ into Eq.
(\ref{swt}) is due to the following relation, valid for integer values of $l$:
\be
P_l (-\cos \theta) = (-1)^l P_l (\cos \theta) .
\ee
Then, we can reshape the contour $C$ into the straight line $\re (l)
= -{1 \over 2}$. Eq. (\ref{swt}) then becomes
\bea
\tilde{\cal C}_E^{(hh)} (\theta,t)
 = &-& \displaystyle\sum_{ \re (\alpha_n) >  -{1 \over 2} }
{ \pi r_n (t) P_{\alpha_n (t)} (-\cos \theta) \over \sin (\pi \alpha_n (t))}
\nonumber \\
&-& {1 \over 2i} \displaystyle\int_{-{1 \over 2} -i\infty}^{-{1 \over 2} 
+i\infty} { A_l (t) P_l (-\cos \theta) \over \sin (\pi l) } dl ,
\label{swt-E}
\eea
where $\alpha_n (t)$ is a pole of $A_l (t)$ in the complex $l$--plane and
$r_n (t)$ is the corresponding residue. We have assumed that $A_l$ vanishes
enough rapidly as $|l| \to \infty$ in the right half--plane, so that the 
contribution from the infinite contour is zero. Eq. (\ref{swt-E}) immediately
leads to the asymptotic behavior of the scattering amplitude in the limit
$s \to \infty$, with a fixed $t$ ($|t| \ll s$). In fact, making use of the
analytic extension (\ref{final-hh}) when continuing the angular variable,
$\theta \to -i\chi$, we derive that
\bea
\lefteqn{
\tilde{\cal C}_M^{(hh)} (\chi, t) =
\tilde{\cal C}_E^{(hh)} (-i\chi, t)
} \nonumber \\
& & = - \displaystyle\sum_{ \re (\alpha_n) >  -{1 \over 2} }
{ \pi r_n (t) P_{\alpha_n (t)} (-\cosh \chi) \over \sin (\pi \alpha_n (t))}
\nonumber \\
& & - {1 \over 2i} \displaystyle\int_{-{1 \over 2} -i\infty}^{-{1 \over 2} 
+i\infty} { A_l (t) P_l (-\cosh \chi) \over \sin (\pi l) } dl .
\label{swt-M}
\eea
The hyperbolic angle $\chi$ is linked to $s$ by the relation (\ref{logs}).
The asymptotic form of $P_\alpha (z)$ when $|z| \to \infty$ is well known.
It is a linear combination of $z^\alpha$ and of $z^{-\alpha -1}$.
When $\re (\alpha) >  -1/2$, this last term can be neglected.
Therefore, in the limit $s \to \infty$, with a fixed $t$ ($|t| \ll s$),
we are left with the following expression:
\be
\tilde{\cal C}_M^{(hh)} (\chi \to \infty, t)
\sim \displaystyle\sum_{ \re (\alpha_n) >  -{1 \over 2} }
\beta_n (t) s^{\alpha_n (t)} .
\label{CMhh-regge}
\ee
The integral in Eq. (\ref{swt-M}), usually called the {\it background term},
vanishes at least as $1/\sqrt{s}$.
Eq. (\ref{CMhh-regge}) allows to immediately extract the scattering
amplitude according to Eq. (\ref{scatt-hadron2}):
\be
{\cal M}_{(hh)} (s,t) = -i~2s~
\tilde{\cal C}_M^{(hh)} (\chi \to \infty, t)
\sim -2i \displaystyle\sum_{ \re (\alpha_n) > -{1 \over 2} }
\beta_n (t) s^{1+\alpha_n (t)} .
\label{Mhh-regge}
\ee
This equation gives the explicit $s$--dependence of the scattering amplitude at 
very high energy ($s \to \infty$) and small transferred momentum ($|t| \ll s$).
As we can see, this amplitude comes out to be a sum of powers of $s$.
This sort of behavior for the scattering amplitude was first proposed by 
Regge \cite{Regge} and $1+\alpha_n (t)$ is often called
a ``{\it Regge pole}''. In the original derivation \cite{Regge}, 
the asymptotic behavior (\ref{Mhh-regge}) was recovered by analytically 
continuing to very large imaginary values the angle between the
trajectories of the two exiting particles in the $t$--channel process.
Instead, in our derivation, we have used the Euclidean--to--Minkowskian
analytic continuation (\ref{final-hh}) and we have analytically continued
the Euclidean correlator to very large (negative) imaginary values 
of the angle $\theta$ between the two Euclidean Wilson loops.
As in the original derivation, we have assumed that the singularities of
$A_l(t)$ in the complex $l$--plane (at a given $t$) are only simple poles
in $l_n = \alpha_n (t)$. If there are other kinds of singularities, 
different from simple poles, their contribution will be of a different type
and, in general, also logarithmic terms (of $s$) may appear in the
amplitude.

Denoting with $\overline{\alpha}(t)$ the pole with the largest real part (at
that given $t$), we thus find that:
\be
\tilde{\cal C}_M^{(hh)} \left( \chi \mathop{\sim}_{s \to \infty}
\log \left( {s \over m_1 m_2} \right), t \right)
\sim \overline{\beta}(t) s^{\overline{\alpha}(t)} .
\label{CMhh-asympt}
\ee
This implies, for the hadron--hadron elastic scattering amplitude
(\ref{Mhh-regge}), the following high--energy behaviour:
\be
{\cal M}_{(hh)} (s,t) = -i~2s~
\tilde{\cal C}_M^{(hh)} (\chi \to \infty, t) 
\sim -i~2\overline{\beta} (t)~ s^{1+\overline{\alpha}(t)} ,
\label{Mhh-asympt}
\ee
and, therefore, by virtue of the optical theorem (\ref{optical}):
\be
\sigma_{\rm tot}^{(hh)} (s) \mathop{\sim}_{s \to \infty}
{1 \over s} {\rm Im} {\cal M}_{(hh)} (s, t=0)
\mathop{\sim} \sigma_0^{(hh)} \left( {s \over s_0} \right)^{\epsilon_P} ,
~~~~{\rm with:}~~\epsilon_P = \re[\overline{\alpha}(0)] .
\label{pomeron2}
\ee
We want to stress two important issues which clarify under which conditions
we have been able to derive this {\it pomeron}--like behaviour for the
elastic amplitudes and the total cross sections.

{\bf i)} We have ignored a possible energy dependence of hadron wave functions
and we have thus ascribed the high--energy behaviour of the Minkowskian
hadronic correlation function exclusively to the {\it fundamental} loop--loop
correlation function (\ref{CMtilde}). With this hypothesis, the coefficients
$A_l$ in the partial--wave expansion (\ref{pwe}) and, as a consequence, the
coefficients $\beta_n$ and $\alpha_n$ in the Regge expansion (\ref{CMhh-regge})
do not depend on $s$, but they only depend on the Mandelstam variable $t$.

{\bf ii)} However, this is not enough to guarantee the experimentally--observed
{\it universality} (i.e., independence on the specific type of hadrons
involved in the reaction) of the {\it pomeron} trajectory
$\overline{\alpha}(t)$ in Eq. (\ref{Mhh-asympt}) and, therefore, of the
{\it pomeron} intercept $\epsilon_P$ in Eq. (\ref{pomeron2}).
In fact, the partial--wave expansion (\ref{pwe})
of the hadronic correlation function can be considered, by virtue of
Eqs. (\ref{CMEhh}) and (\ref{CMEhh-tilde}), as a result of a partial--wave
expansion of the {\it fundamental} loop--loop Euclidean correlation function
(\ref{CEtilde}), i.e.,
\be
\tilde{\cal C}_E (\theta,t;~\vec{R}_{1\perp},\vec{R}_{2\perp}) =
\displaystyle\sum_{l=0}^{\infty}
{\cal A}_l(t;~\vec{R}_{1\perp},\vec{R}_{2\perp}) P_l (\cos \theta) ,
\label{CEtilde-pwe}
\ee
which is then averaged with two proper squared hadron wave functions:
\be
\tilde{\cal C}_E^{(hh)} (\theta,t) =
\displaystyle\int d^2\vec{R}_{1\perp} |\psi_1(\vec{R}_{1\perp})|^2
\displaystyle\int d^2\vec{R}_{2\perp} |\psi_2(\vec{R}_{2\perp})|^2
\nonumber \\
\tilde{\cal C}_E (\theta,t;~\vec{R}_{1\perp},\vec{R}_{2\perp}) .
\label{CEhh-tilde}
\ee
If we now repeat for the partial--wave expansion (\ref{CEtilde-pwe}) the
same manipulations that have led us from Eq. (\ref{pwe}) to Eq.
(\ref{CMhh-regge}), we arrive at the following Regge expansion for the
loop--loop Minkowskian correlator:
\be
\tilde{\cal C}_M (\chi \to \infty, t;~\vec{R}_{1\perp},\vec{R}_{2\perp})
\sim \displaystyle\sum_{ \re (a_n) >  -{1 \over 2} }
b_n (t;~\vec{R}_{1\perp},\vec{R}_{2\perp})
s^{a_n(t;~\vec{R}_{1\perp},\vec{R}_{2\perp})} ,
\label{CM-regge}
\ee
where $a_n(t;~\vec{R}_{1\perp},\vec{R}_{2\perp})$ is a pole of
${\cal A}_l (t;~\vec{R}_{1\perp},\vec{R}_{2\perp})$ in the complex $l$--plane.
After inserting the expansion (\ref{CM-regge}) into the expression for the
Minkowskian hadronic correlation function, i.e.,
\be
\tilde{\cal C}_M^{(hh)} (\chi,t) =
\displaystyle\int d^2\vec{R}_{1\perp} |\psi_1(\vec{R}_{1\perp})|^2
\displaystyle\int d^2\vec{R}_{2\perp} |\psi_2(\vec{R}_{2\perp})|^2
\tilde{\cal C}_M (\chi,t;~\vec{R}_{1\perp},\vec{R}_{2\perp}) ,
\label{CMhh-tilde}
\ee
one in general finds a high--energy behaviour which hardly fits with that
reported in Eqs. (\ref{CMhh-asympt}) and (\ref{Mhh-asympt}) with a
{\it universal} {\it pomeron} trajectory $\overline{\alpha}(t)$, {\it unless}
one assumes that, for each given loop--loop correlation function with
transverse separations $\vec{R}_{1\perp}$ and $\vec{R}_{2\perp}$, (at least)
the position of the pole $a_n(t;~\vec{R}_{1\perp},\vec{R}_{2\perp})$ with the
largest real part does not depend on $\vec{R}_{1\perp}$ and $\vec{R}_{2\perp}$,
but only depends on $t$. (Maybe this is a rather {\it natural} assumption if
one believes that the {\it pomeron} trajectory is, after all, determined by an
even more fundamental quantity, that is the line--line, i.e., parton--parton,
correlation function.)
If we denote this ``common'' pole with $\overline{\alpha}(t)$,
we then immediatelly recover the high--energy behaviour (\ref{CMhh-asympt}),
where the coefficient in front is given by:
\be
\overline{\beta}(t) =
\displaystyle\int d^2\vec{R}_{1\perp} |\psi_1(\vec{R}_{1\perp})|^2
\displaystyle\int d^2\vec{R}_{2\perp} |\psi_2(\vec{R}_{2\perp})|^2
b_n (t;~\vec{R}_{1\perp},\vec{R}_{2\perp}) ,
\label{beta-tilde}
\ee
and therefore, differently from the {\it universal} function
$\overline{\alpha}(t)$, explicitly depends on the specific type of
hadrons involved in the process.\\

\noindent
In conclusion, we have shown that the Euclidean--to--Minkowskian
analytic--continuation approach can, with the inclusion of some extra
(more or less plausible) assumptions, easily reproduce a {\it pomeron}--like
behaviour for the high--energy total cross sections.
However, we should also keep in mind that the {\it pomeron}--like behaviour
(\ref{pomeron}) is, strictly speaking, forbidden (at least if considered as a
true {\it asymptotic} behaviour) by the well--known Froissart--Lukaszuk--Martin
(FLM) theorem \cite{FLM} (see also \cite{Heisenberg}), according to which,
for $s \to \infty$:
\be
\sigma_{\rm tot}(s) \le {\pi \over m_\pi^2} \log^2 \left( {s \over s_0}
\right) ,
\label{FLM}
\ee
where $m_\pi$ is the pion mass and $s_0$ is an unspecified squared mass
scale.\\
In this respect, the {\it pomeron}--like behaviour (\ref{pomeron}) can at most
be regared as a sort of {\it pre--asymptotic} (but not really
{\it asymptotic}!) behaviour of the high--energy total cross sections
(see, e.g., Refs. \cite{BB,Kaidalov} and references therein).
Immediately the following question arises: why our approach, which was
formulated so to give the really asymptotic large--$s$ behaviour of
scattering amplitudes and total cross sections, is also able to reproduce
pre--asymptotic behaviours [violating the FLM bound (\ref{FLM})] like the
one in (\ref{pomeron})?
The answer is clearly that the {\it extra assumptions}, i.e., the {\it models},
which one implicitly or explicitly assumes in the calculation of the
Euclidean correlation functions $\tilde{\cal C}_E$ play a fundamental
role in this respect. This is surely a crucial point which, in our opinion,
should be further investigated (and maybe also better formulated) in the
future. A great help could be provided by a direct lattice calculation
of the loop--loop Euclidean correlation functions. Clearly a lattice approach
can at most give (after having overcome a lot of technical difficulties)
only a discrete set of $\theta$--values for the above--mentioned functions,
from which it is clearly impossible (without some extra assumption on the
interpolating continuous functions) to get, by the analytic continuation
$\theta \to -i \chi$, the corresponding Minkowskian correlation functions
(and, from this, the elastic scattering amplitudes and the total cross
sections). However, the lattice approach could provide a criterion to
investigate the goodness of a given existing analytic model (such as:
Instantons, SVM, AdS/CFT, BFKL and so on $\ldots$) or even to open the way
to some new model, simply by trying to fit the lattice data with the
considered model.\\
This would surely result in a considerable progress along this line
of research.


\end{document}